\documentclass[twocolumn,amsmath,amssymb,eqsecnum,floatfix,rmp]{revtex4}

\usepackage{graphicx}
\usepackage{dcolumn}
\usepackage{bm}
\usepackage{url}

\newcommand\eq{\begin{equation}}
\newcommand\en{\end{equation}\noindent}

\newcommand\cH{{\cal H}}
\newcommand\cL{{\cal L}}
\newcommand\wwp{{\rho^{\prime}}}
\newcommand\wdp{{\rho^{\prime\prime}}}
\newcommand\rp{{r^{\prime}}}
\newcommand\rdp{{r^{\prime\prime}}}
\newcommand\zp{{z^{\prime}}}
\newcommand\zdp{{z^{\prime\prime}}}
\newcommand\ztp{{z^{\prime\prime\prime}}}
\newcommand\zqp{{z^{\prime\prime\prime\prime}}}

\newcommand\mflat{\mu_{{\text{flat}}}}
\newcommand\csat{c_{{\text{sat}}}}
\newcommand\msur{\mu_{{\text{sur}}}}
\newcommand\msol{\mu_{{\text{sol}}}}
\newcommand\bx{\textbf{x}}
\newcommand\bn{\textbf{n}}

\newcommand\wide{\begin{widetext}}
\newcommand\wideend{\end{widetext}}

\begin{document}   
\def\I.#1{\it #1}   
\def\B.#1{{\bbox#1}}  
\def\C.#1{{\cal #1}}
\def\vx {\vec{x}}
\def\vk {\vec{k}}
\def\df {\delta \phi}
\def\dm {\delta m}
\def\pl {\partial}
\def\X  {{\bf{X}}}
\def\N  {{\bf{N}}}
\def\tiz {{\tilde{z}}}
\def\ttr {{\tilde{\tilde{r}}}}
\def\ttz {{\tilde{\tilde{z}}}}

\title{Reaction-limited sintering in nearly saturated environments}
\author{Benny Davidovitch\footnote{Current address: Division of Engineering 
and Applied Sciences, Harvard University, Cambridge, MA 02138}
, Deniz Erta{\c s}, and Thomas C. Halsey}  
\affiliation{Corporate Strategic Research, ExxonMobil Research and Engineering, 
Route 22, Annandale, NJ 08801} 

\begin{abstract}
We study 
the shape and growth rate of necks between sintered spheres
with dissolution-precipitation dynamics in the reaction-limited regime.
We determine the critical shape that separates those  
initial neck shapes that can sinter 
from those that necessarily dissolve, as well as the 
asymptotic evolving shape
of sinters far from the critical shape.
We compare our results with past results for 
the asymptotic neck shape in closely related but more complicated 
models of surface dynamics; in particular we confirm a scaling conjecture,
originally due to Kuczinsky.
Finally, we consider the relevance of this problem to the diagenesis of
sedimentary rocks and other applications. 
\end{abstract}

\maketitle
\section{Introduction}
Sintering is a surface-tension driven phenomenon in which solid 
particles packed below their melting temperature are consolidated
via the growth of 
necks and subsequent shrinkage of pores between the particles.
Quantitative understanding of sintering is crucial to study 
evolution of porous morphologies in industry and in nature -- key 
properties of ceramic and metallic powders, 
like their strength and degree of compactness,
are controlled by the sintering process \cite{kuczinsky49,herring50,kingery55,german96}. 
Packed snow flakes and ice spheres may be bonded together
by necks well below their melting temperature -- 
a phenomenon that is important for understanding 
snow avalanches and glacier flows \cite{kingery60,kuroiwa61,hobbs64,maono83,colbeck98}. 
Sedimentary and other granular rocks also undergo sintering 
processes that affect their porosity and permeability \cite{jurewicz85,hay88,visser99}. 

Sintering processes typically exhibit
several stages, identified by the degree of
porosity of the material and the nature of the dominant 
surface tension, as well as the kinetic route. In the early 
stage of sintering, when particles are barely touching each other, 
the driving force is the surface tension between the solid 
particles and the surrounding pore space -- be it filled with vapor, 
solution, or vacuum. 
The chemical potential $\msur ({\bf x})$ associated with surface 
tension at a point ${\bf x}$ on the surface is given by
$\msur ({\bf x}) = \sigma \nu_m H ({\bf x})$,
where $\nu_m$ is a molecular volume in the solid phase,
$H = (\kappa_1  + \kappa_2 )/2$ is the mean between the two principal 
curvatures $\kappa_{1,2}$, and $\sigma$ is the surface tension, here taken to be isotropic for simplicity. 
Typically, the contact zone between the particles is 
highly concave and has a negative $H$, 
whereas the particles themselves are convex with positive $H$. 
This difference in $H$, and hence in the chemical potential 
$\msur ({\bf x})$, gives rise to a net flow
of solid molecules from the periphery to the touching zone and to the 
formation and growth of necks between the particles.

At later stages of the process, necks become thicker and
variations in $H$ along the surface are diminished.
Most of the pores are now concentrated inside grains and on
grain boundaries, and the thermodynamic driving force for 
further compactification is not $\sigma$, 
but rather the grain-boundary energy. 
The dominant kinetic mechanisms in these stages
are associated mainly with bulk transport: 
grain-boundary diffusion, plastic flow and lattice diffusion \cite{swinkels81,german86}. 

Here we focus on early stage sintering where the 
mass transfer is controlled solely by 
dissolution-precipitation through a surrounding solution. 
Such kinetics is dominant when the solid phase is in coexistence with 
vapor or solution, and the temperature
is not too close to the melting temperature, at which 
thermal energy is high enough to activate fast surface diffusion
processes. In addition, we assume that thermodynamics (energy) and 
kinetics (diffusion) associated with grain boundaries can be neglected, as 
is the case, e.g.,  with amorphous materials.  
We will limit ourselves to cases
in which the chemical reaction rate for dissolution-precipitation 
between the solid and solution is much slower than 
typical rates of molecular transport through the surrounding 
media. Under such reaction-limited dynamics 
spatial variations of $\msol$, the chemical potential
in the solution, can be neglected. 
Moreover, we will consider particles in solution under `open' conditions,
in which the total amount of solid material is not conserved,
but the solute concentration, and hence $\msol$, is fixed in time. 
These assumptions are unrealistic for 
industrial sintering of powders, in which 
proper inclusion of molecular transport and other considerations can be 
crucial.

The prototypical geometry for early stage sintering, introduced
by Kuczinsky \cite{kuczinsky49}, and studied extensively since then, 
is of a cylindrical neck between two touching spherical 
particles. 
Assuming a characteristic shape for the 
evolving neck profile, it was shown \cite{kuczinsky49,herring50},  
that for reaction-limited dynamics, dominated solely by 
dissolution-precipitation,    
the neck thickens in time as $t^{1/3}$ at short times. 
If diffusion through the vapor is much slower than the reaction rate, 
it was argued that the neck thickens as $t^{1/5}$ \cite{hobbs64}. 
Other kinetic routes give rise as well to various power laws.   
During the last 50 years, many experiments aimed to test these theories, 
and power law growth rates of necks were indeed 
found \cite{kingery55,kingery60,hobbs64}. 
For sintering dominated by surface diffusion,
several researchers argued, based on numerical simulations and
analytic arguments, that Kuczinsky's growth rates are not correct
(Nichols and Mullins 1965, German and Lathrop 1978, Eggers 1998).
However, the multitude of microscopic mechanisms for sintering, and the 
fact that in real physical processes several of them may be significant, 
makes the verification of a simple theory a challenging task for
the experimentalist. 

In this paper we present a detailed
study of sintering driven by our model dynamics in this geometry.
We focus on two aspects of the surface dynamics:
the critical surface, which is actually an unstable
fixed point of the appropriate surface dynamics, and 
the asymptotic growth profile.   
Our analysis of the critical surface provides us with conditions
on initial contact geometries that can become persistent sinters, and 
enables us to determine 
growth rates at very early stages of the sintering process.
For the asymptotic profile, we verify the $t^{1/3}$ 
growth rate for the neck, derive analytically the correct 
neck profile, and 
explain how it selects the appropriate growth rate.  

The paper is organized as follows:
In section 2 we introduce our model system and derive an equation of motion 
for cylindrically symmetric surface evolution. 
In section 3 we study the structure of the unstable 
fixed point of this equation
(the critical surface), 
and perform linear stability analysis of its dynamics. 
In section 4 we discuss the evolution of surfaces far from the critical
surface, and show how the asymptotic dynamics can be described as 
a self-consistent solution of the equation of motion.
In section 5 we conclude and suggest future directions. An appendix contains 
some calculational details.

\section{Equation of surface evolution}
Let us consider a solid in coexistence with its 
solution in a surrounding liquid (or vapor), and denote the interface 
between the two phases as $\Gamma \equiv \bx(u,v)$. 
We assume a first order single-component chemical reaction between the 
solid and solution, characterized by $K_f$ - the dissolution rate of 
a flat solid interface to a `fresh' (unsaturated) solution.
We assume that the surface tension $\sigma$ between
the two phases can be considered as isotropic (which will be true
well above the roughening transition or for an amorphous solid).
The normal velocity of the surface $u_n (\bx)$ into the fluid region 
is given by the difference between dissolution and precipitation rates
\begin{equation}
u_n (\bx) = -K_f \left( 1 - e^{-\frac{\Delta \mu (\bx) }{kT}} \right)~, 
\label{surfacedynamics1}
\end{equation}
where the precipitation rate is controlled by the Boltzmann factor 
associated with the difference 
$\Delta \mu (\bx) \equiv  \msur (\bx) - \msol (\bx)$
between the chemical potentials of solid and dissolved molecules on the
two sides of the interface $\Gamma$. 
The chemical potentials $\msur (\bx)$ and $\msol (\bx)$ are given by  
\begin{eqnarray}
\msur (\bx) &=& 2 \nu_m \sigma H(\bx) + \mflat  ~, 
\label{msurf} 
\\
\msol (\bx) &=& kT \log \frac{c(\bx)}{\csat} + \mflat ~,
\label{satchem}
\end{eqnarray} 
where $\mflat$ is the chemical potential of a flat surface in equilibrium
with a saturated solution at concentration $\csat$,
$c({\bf x})$ is the concentration near the surface point $\bf{x}$,
$\nu_m$ is the molecular volume in the solid,
and an ideal solution is assumed.
Here and elsewhere we define all concentrations with respect to the 
concentration in the solid. Equation (\ref{msurf}) follows 
from the geometrical identity $H(\bx)= (1/2)(\delta S/\delta V)$,
relating the mean curvature and the local variation in surface area 
$\delta S$ with respect to a volume change $\delta V$ of the solid. 
For our system we assume that the surface tension energy is much smaller 
than $kT$, and that the solution is nearly saturated. Equation 
(\ref{surfacedynamics1}) then reduces to: 
\begin{equation}
u_n (\bx) =
-K_f \frac{2\nu_m \sigma}{kT} \left [ H (\bx) - \frac{kT}{2\nu_m \sigma}
\log \frac{c(\bx,t)}{\csat} \right ].
\label{dissol}
\end{equation} 
In order to fully describe the evolution of a solid-fluid 
interface $\Gamma (t)$, 
equation (\ref{dissol}) must be supplemented by an advection-diffusion 
equation governing the concentration in the 
solution, for which $c({\bf x}, t)$ is the boundary value at the interface with the solid phase. 

In this study we are interested in kinetic conditions for which any 
inhomogeneity in the concentration $c({\bf x},t)$ equilibrates 
much faster than the typical time for chemical 
reactions to cause significant changes of the surface shape by 
dissolution-precipitation events. In other words, we assume:
\begin{equation} 
\tau_{{\rm diss}} \gg \tau_{{\rm diff}} \qquad {\rm{or}} \qquad
\tau_{{\rm diss}} \gg \tau_{{\rm adv}}, 
\label{typicaltimes1}
\end{equation}
where the kinetic time scales are defined by
\begin{equation}
\tau_{{\rm diss}} = \frac{L_{{\rm par}}}{K_f}, \qquad 
\tau_{{\rm diff}} = \frac{L_{{\rm pore}}^2}{D}, \qquad 
\tau_{{\rm adv}} = \frac{L_{{\rm pore}}}{V},
\label{connecttimes}
\end{equation} 
where $D$ is the diffusion constant, $V$ is typical velocity of the 
advecting flow, $L_{{\rm par}}$ is a typical size of a particle, and 
$L_{{\rm pore}}$ is a typical separation between particles.
Such kinetic conditions are typically realized in sedimentary rocks 
that dissolve extremely slowly, 
such that their surface morphology evolves over geological time 
scales. 
     
Under this reaction-limited dynamics, we can assume 
a homogeneous distribution of the concentration and 
hence the chemical potential in the solution:
$c({\bf x},t) \to c_{\infty}(t)$, fixed by conditions outside 
the local region of interest.
The surface dynamics equation (\ref{dissol}) then reduces to:
\begin{eqnarray}
u_n (\bx) &=&
-K_f \frac{2\nu_m \sigma}{kT} \left [ H (\bx) - H_* \right ] ,
\label{surdyn}
\\
H_* &\equiv& [{kT}/({2\nu_m \sigma})]\log (c_{\infty}(t) / \csat) .  
\label{defineHstar}
\end{eqnarray} 
For $H_*=0$ equation (\ref{surdyn}) becomes the well-known Allen-Cahn 
equation, describing the decay to global equilibrium of a binary system
under the influence of surface tension \cite{bray94}.

In this paper we focus on the evolution under (\ref{surdyn}) of a sinter 
between spheres of radius $R$. 
Typical curvatures in the neck region 
are much higher (in absolute value) than the curvature of the spheres, 
and therefore we take $c_{\infty}(t)= \csat \exp(2 \nu_m \sigma/kT R)$, which guarantees that before contact is made 
the two solid spheres are in equilibrium with the solution.     
This condition replaces the more common condition 
of conservation of solid matter, which is usually encountered in 
applications in material science, and which can be implemented 
in the open, reaction-limited case by suitably varying $H_*$ (slowly) with time. 

Depending on the surface geometry in the contact region, the dynamics
(\ref{surdyn}) may lead to dissolution of the contact, or to sintering - 
a neck growing between the two spheres.
Evolution of surfaces under equation (\ref{surdyn}) 
preserves cylindrical symmetry with respect to the axis 
connecting the centers of the two spheres.
\begin{center}
\begin{figure}
\includegraphics[scale=0.5]{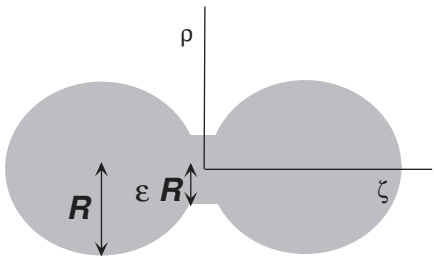}
\caption{Two sintering spheres--the sinter preserves the
cylindrical symmetry around the $\zeta$-axis. The width of the neck divided by 
the diameter of the sphere defines the small parameter $\epsilon$.}
\label{figure1}
\end{figure}
\end{center}

Let us rewrite equation~(\ref{surdyn}), taking advantage of this cylindrical 
symmetry. Defining the $\zeta$-axis as the axis of this symmetry, with $\rho(\zeta,t)$ 
the radial position of the surface (see Fig.~\ref{figure1}), 
the in-plane (longitudinal) curvature $\kappa_l$ and the out-of-plane (azimuthal) 
curvature $\kappa_a$ are:
\eq
\kappa_l = -\frac{\wdp}{[1+\wwp^2]^{3/2}}, \qquad
\kappa_a = \frac{1}{\rho\sqrt{1+\wwp^2}},
\en
yielding a mean curvature $H$ of
\eq
H = \frac{1}{2} (\kappa_a  + \kappa_l) = 
\frac{1}{2\sqrt{1+\wwp^2}} \left(\frac{1}{\rho} - \frac{\wdp}{1+\wwp^2} \right) .
\label{cscurv}
\en
Furthermore, the normal velocity $u_n$ is given by:
\eq
u_n = (1+\wwp^2)^{-1/2}\partial \rho/\partial t.
\en
Upon rescaling time and length by:
\begin{equation}
t \to \frac{t}{kT / ( R^2 K_f   \nu_m \sigma )}, \qquad 
 r = \rho/R , \qquad z = \zeta / R,
\label{rescale}
\en 
Equation~(\ref{surdyn}) takes the form:
\eq
\frac{\partial r}{\partial t} = -\left[\frac{1}{r} - 
\frac{\rdp}
{1+\rp^2}- 
2\sqrt{1 + \rp^2}   
\right]  . 
\label{cylsurdyn2} 
\en    
In the following sections we will study the surface 
dynamics described by equation 
(\ref{cylsurdyn2}).

\section{Fixed points}
In this section we will discuss equation~(\ref{cylsurdyn2}) near its fixed
points - surfaces for which ${\partial r}/{\partial t} =0$ at each point.
From equations (\ref{surdyn},\ref{cscurv}) 
we see that such surfaces are cylindrically 
symmetric surfaces of constant mean curvature (CSCMC)
whose mean curvature $H(z) = 1$. 
First we discuss the structure of these surfaces, then we use linear stability 
analysis to study 
the dynamics near such surfaces.

\subsection{Structure}
According to equation (\ref{cylsurdyn2}), CSCMC are defined by solutions
to the equation:
\eq
 2 \sqrt{1+{\rp}^2} = \frac{1}{r} - 
\frac{\rdp}{1+{\rp}^2}  . 
\label{cylsurfp} 
\en
This is a second order ordinary differential equation (ODE), and therefore
two boundary conditions (BC) are required for a unique solution.
Since we are interested here in sintering of two identical spheres,
we consider the boundary conditions to be:
\eq
r(z=0) =\epsilon, \qquad r^{\prime}(z=0) = 0, 
\label{cylsurfpbc}
\en
where $z=0$ is the middle point between the centers of the touching spheres,
$\epsilon$ is the ratio between the neck size and the radius $R$ of the 
unsintered spheres (which determines $H_*$), and the latter condition 
reflects the symmetry between the two sides of the contact.
\begin{center}
\begin{figure}
\includegraphics[scale=0.45]{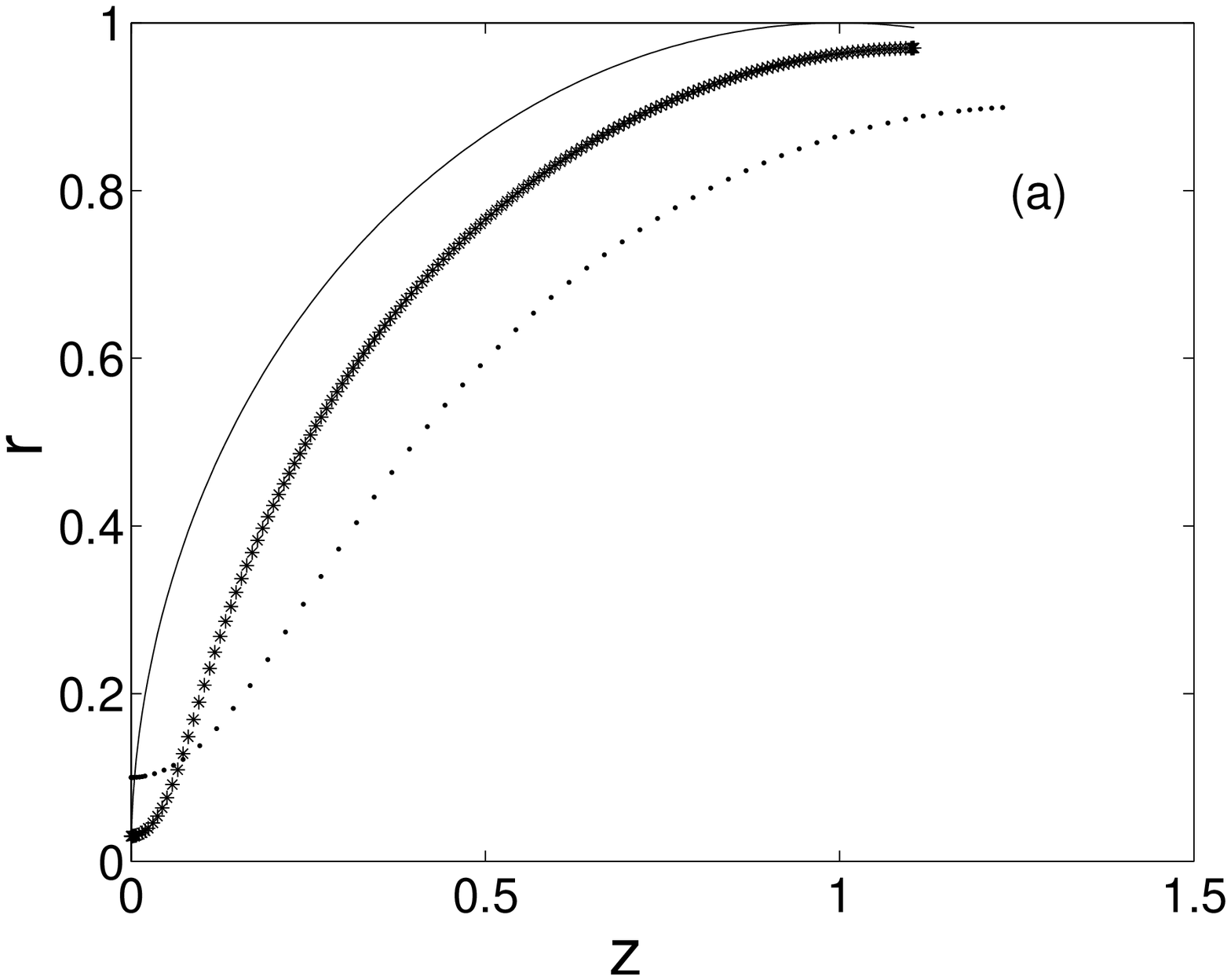}
\includegraphics[scale=0.45]{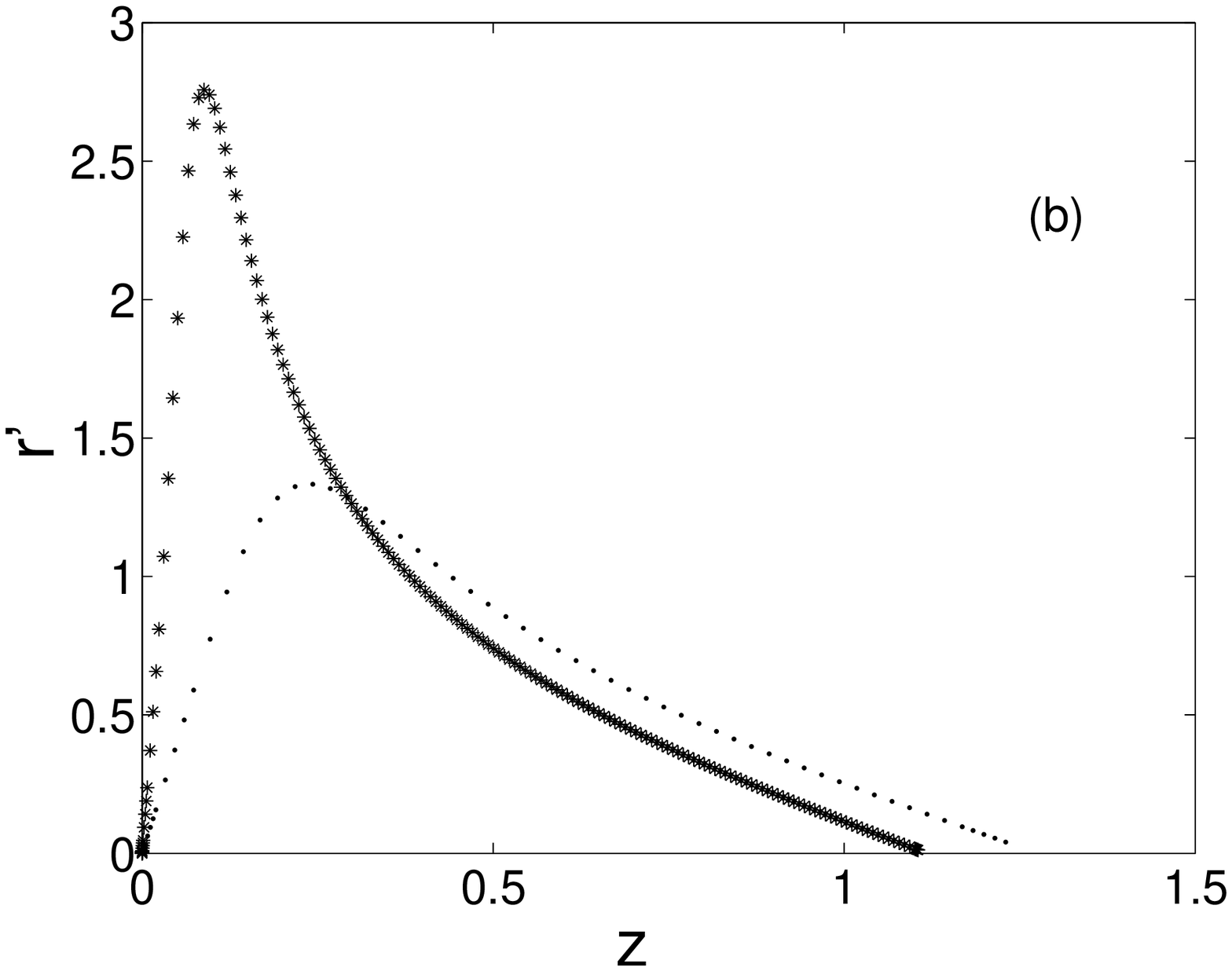}
\caption{Profile of a cylindrically symmetric constant mean curvature surface,
obtained by numerical integration of equations~(\ref{cylsurfp},\ref{cylsurfpbc}), for 
two values of $\epsilon$: $\epsilon=0.1$ (dots) , and $\epsilon = 0.03$ (stars):
(a) The profile $r(z)$. The solid curve is a spherical profile $r_{sph}(z)$. 
(b) The derivative of the profile $\rp(z)$.}
\label{figure2}
\end{figure}
\end{center}

Equations~(\ref{cylsurfp},\ref{cylsurfpbc}) can be easily integrated 
numerically. The profiles $r(z)$ and their derivatives $\rp(z)$ for two
values of $\epsilon$ are shown in figure 2, together with the profile 
$r_{sph}(z)$ of a sphere of radius $1$, whose center lies on the $z$-axis 
at $z=1$. Obviously, $r_{sph}^{\prime}(z)$ diverges as $z \to 0$.

The appearance of a singular inner region near $z=0$ in which 
$|r^{\prime} (z) - r_{sph}^{\prime}(z)| >>  O(\epsilon) $ 
indicates that a regular perturbative solution 
to equation~(\ref{cylsurfp}) 
is not possible; this equation therefore 
calls for singular perturbation analysis. 

\subsubsection{Singular perturbation analysis}   

Another form of equation~(\ref{cylsurfp}) is achieved by considering $z$ as a 
function of $r$ (assuming that the function $r(z)$ is one-to-one):
\eq
2 \sqrt{1+\zp^2} = \frac{\zp}{r} + 
\frac{\zdp}{1+\zp^2} \ ,
\label{cylsurfpz}
\en
and the transformation of the BC~(\ref{cylsurfpbc}) is:
\eq
z(r=\epsilon) = 0, \qquad
\zp(r \to \epsilon^{+}) \to  \infty .  
\label{cylsurfpbcz}
\en 
Equations~(\ref{cylsurfpz},\ref{cylsurfpbcz})
are more convenient for analytic study and will be used in the sequel  
instead of equations~(\ref{cylsurfp},\ref{cylsurfpbc}).

In order to construct a series expansion in $\epsilon$ for  
$z (r)$ in the interval $r \in [\epsilon,1]$, one needs to   
consider separately the neck and the periphery, identify the small 
non-dimensional parameters in each of them, and then match the  
two expansions. 

Let us start with the peripheral region far from the neck, 
which we call the `outer' region: $r \gg \epsilon$.
In this region 
we can expand the profile $z_{out}(r)$ as a series 
\eq
z_{out}(r) = z^{(0)}_{out} (r) + \epsilon z^{(1)}_{out} (r) +  
O(\epsilon ^2) , 
\label{outerexpand}
\en
where $z^{(0)}_{out}(r)$ is the profile of a sphere of radius 
$1$, centered on the $z$-axis at $(r=0 , z=1)$:
\eq
{z^{(0)}}_{out} (r) =  1- \sqrt{1 -r^2} \ . 
\label{outer0} 
\en

We now consider the next order in $\epsilon$. Substituting in 
(\ref{cylsurfpz}) the expansion~(\ref{outerexpand}) and
the function $z_{out}^{(0)}$ (\ref{outer0}), we 
obtain the following equation for $z_{out}^{(1)}$:
\eq
\frac{4{r^2 - 1}}{r(1-r^2)} 
(z_{out}^{(1)})^{\prime} = 
(z_{out}^{(1)})^{\prime\prime}
\en
whose solution satisfies
 
\eq
(z^{(1)}_{out})^{\prime} = \frac{a}{r(1-r^2)^{3/2}} , 
\label{houteps} 
\en 
where $a$ is a constant of integration to be ultimately determined by a matching
condition.
Thus, to first order in $\epsilon$, the derivative of the outer 
solution is:
\eq
z_{out}^{\prime} = \frac{r}{\sqrt{1-r^2}} + \epsilon \frac{a}{r (1-r^2)^{3/2}} + 
O(\epsilon^2)  . 
\label{uptofirstout}
\en

A necessary condition for equation~(\ref{outerexpand}) 
to be a valid asymptotic expansion is that: 
$\vert (z_{out}^{(1)})^{\prime} \vert \ll \vert (z_{out}^{(0)})^{\prime} \vert$. This 
leads to 
the conditions: 
\begin{equation}
r \gg \sqrt{\epsilon}, \qquad  1-r \gg \sqrt{\epsilon} .   
\label{assumout}
\en
Condition~(\ref{assumout}) indicates the existence of an `equatorial'
region $1 - \sqrt{\epsilon} < r  <1$, where the outer perturbative 
expansion~(\ref{outerexpand}) is not valid, in which the profile is described 
by a different function $z_{equ}(r)$ that matches $z_{out}(r)$. As we shall
see below, this actually amounts to a displacement of the singularity 
at $r \to r_s=1$ in equations~(\ref{outer0},\ref{houteps}) 
to $r \to r_s<1$.  
Since our primary interest here is the neck structure, we postpone 
the analysis of $z_{equ}$ to the end of this section.    

The divergence of $\zp$ required
by condition (\ref{cylsurfpbcz}) indicates that a regular 
expansion like~(\ref{outerexpand})
cannot satisfy equation~(\ref{cylsurfpz}) subject to the BC~(\ref{cylsurfpbcz}) 
\cite{hinch91}.
This fact implies that a perturbative solution 
to equations~(\ref{cylsurfpz},\ref{cylsurfpbcz}) must be constructed from an expansion 
whose $0^{th}$ order term is different from (\ref{outer0}). 
We call this expansion $z_{in}(z)$,
and write, similarly to equation~(\ref{outerexpand}):
\eq
z_{in}(r) = z^{(0)}_{in} (r) + \epsilon z^{(1)}_{in} (r) +  
O(\epsilon ^2) \ .
\label{innerexpand}
\en
The function $z_{in} (r)$ is required to satisfy 
equation~(\ref{cylsurfpz}) under the  
BC~(\ref{cylsurfpbcz}), and therefore we expect it to describe the profile
near the contact between the two spheres.

The BC~(\ref{cylsurfpbcz}) indicates that 
a natural length scale at the neck region is $\epsilon$, and therefore
we re-scale,
\eq
\tilde{r} = \frac{r}{\epsilon} \ , \ \tilde{z} = \frac{z_{in}}{\epsilon} \ . 
\label{innervariables}
\en
Equation~(\ref{cylsurfpz}) then becomes:
\begin{equation}
2 \epsilon \sqrt{1 + (\tiz^{\prime})^2} 
= \frac{\tiz^{\prime}}{\tilde{r}}
 +  \frac{\tiz^{\prime\prime}}
{1+(\tiz^{\prime})^2} .
\label{cylsurfpinz}
\end{equation} 
Assuming 
\eq
\epsilon \sqrt{1 + (\tiz^{\prime})^{2}} \ll 
\left\vert \frac{\tiz^{\prime}}{\tilde{r}} \right\vert,
\left\vert \frac{\tiz^{\prime\prime}}{1+(\tiz^{\prime})^{2}} \right\vert   , 
\label{assum0}
\en
and the expansion (\ref{innerexpand}),
we see that ${\tilde{z}}^{(0)} = z^{(0)}_{in} / \epsilon$ satisfies
the equation:
\eq
\frac{(\tiz^{(0)})^{\prime}} {{\tilde{r}}} +  
\frac{(\tiz^{(0)})^{\prime\prime}} {1+{(\tiz^{(0)})^{\prime}}^2} 
= 0  , 
\label{cylsurfpin0z}
\en 
which is just the equation of a catenoid (CSCMC with zero mean curvature).
The solution of equation~(\ref{cylsurfpin0z}) subject to the 
BC~(\ref{cylsurfpbcz}) satisfies:
\eq
({\tilde{z}}^{(0)})^{\prime} = \frac{1}{\sqrt{{\tilde{r}}^2 -1}} \ . 
\label{innercatenoid}
\en
Now let us calculate the first order term $z^{(1)}_{in}$ 
of the inner expansion.
Substituting in equation (\ref{cylsurfpinz}) the expansion (\ref{innerexpand}) and 
the catenoid function $\tiz^{(0)}$, we get the following equation
for $\tiz^{(1)}= z^{(1)}_{in}/\epsilon$:
\eq
2\frac{\tilde{r}}{\sqrt{{\tilde{r}}^2-1}} = 
\left(\frac{1}{{\tilde{r}}} + \frac{2}{{\tilde{r}}^3}\right) (\tiz^{(1)})^{\prime} + 
\frac{{\tilde{r}}^2-1}{{\tilde{r}}^2}  (\tiz^{(1)})^{\prime\prime} ,
\en
whose solution satisfies:
\eq
(\tiz^{(1)})^{\prime} = \frac{{\tilde{r}}^2}{\sqrt{{\tilde{r}^2 -1}}} .  
\label{hineps} 
\en
The necessary condition 
$\vert (\tiz^{(1)})^{\prime} \vert \ll \vert (\tiz^{(0)})^{\prime} \vert$ implies 
$\tilde r \ll 1/\sqrt{\epsilon}$, or
\eq
r \ll \sqrt{\epsilon}.
\label{validregion0}
\en
which also establishes the validity of equation~(\ref{assum0}).

To first order in $\epsilon$, the inner expansion is thus
\eq
{\tiz}^{\prime} = \frac{1}{{\sqrt{{\tilde{r}}^2 -1}}} + 
\epsilon \frac{{\tilde{r}}^2}{\sqrt{{\tilde{r}}^2-1}} + 
O(\epsilon^2) . 
\label{uptofirstinner}
\en

From equations~(\ref{assumout}) and (\ref{validregion0}) we see that the 
inner and outer expansions are valid in non-overlapping regions in the 
interval $[\epsilon, 1]$. In order to match the 
two expansions $z_{in}$ and $z_{out}$, one has to 
look for a separate solution $z_{mid}$, valid in a `middle' 
region around $r \sim \sqrt{\epsilon}$.
Since the inner and outer expansions are valid at $r \ll \sqrt{\epsilon}$ and 
$r \gg \sqrt{\epsilon}$, respectively, a natural scaling 
for the variable and the function $z_{mid}$ in the middle region is
\eq
\ttr = \frac{r}{\sqrt{\epsilon}} , \qquad  
\ttz = \frac{z_{mid}}{\sqrt{\epsilon}}   .
\en  
The function $z_{mid}$ must match, to first order in $\epsilon$, both
$z_{in}$ and $z_{out}$. This condition, 
and equations~(\ref{uptofirstout},\ref{uptofirstinner}), imply
the following form for the derivative $\ttz^{\prime}$:
\eq
\ttz^{\prime} ( \ttr ) =  a \ttr + \ttr^{-1} + f(\ttr)  
+ O(\epsilon^2)  , 
\label{midregion}
\en   
where asymptotically $f(\ttr)$ must satisfy:
\begin{eqnarray}
\vert f(\ttr) \vert  \ll \ttr^{-1}  \qquad &\rm{at}& \qquad \ttr \ll 1 \nonumber \\ 
\vert f(\ttr) \vert \ll  a \ttr  \qquad &\rm{at}& \qquad \ttr \gg 1  ,  
\label{midbc}
\end{eqnarray}
in order to enable matching to $z_{in}$ and $z_{out}$.
Substituting the form (\ref{midregion}) 
in equation (\ref{cylsurfpz}) and expanding to $O(\epsilon)$ we arrive at the 
solution for $f(\ttr)$: 
\begin{equation}
f(\ttr) = \frac{1}{2} \epsilon (\ttr^3 + \ttr^{-3} + 3) .   
\label{fsol}
\end{equation}
From equations~(\ref{midbc},\ref{fsol}) we see that the asymptotic 
expansion~(\ref{midregion}) for $z_{mid}$ is valid in the region
\begin{equation}
\sqrt{\epsilon} \ll \ttr \ll {\sqrt{\epsilon}^{-1}} , \qquad
\rm{or} \qquad \epsilon \ll r \ll 1 .   
\label{assummid}
\end{equation}  
From equations~(\ref{uptofirstout},\ref{assumout},\ref{validregion0},\ref{uptofirstinner},\ref{midregion},\ref{fsol},\ref{assummid}) we conclude that to first order in $\epsilon$,
$z_{mid}(r)$ and 
$z_{in}(r)$ match in the region $\epsilon \ll r \ll \sqrt{\epsilon}$, 
whereas $z_{mid}(r)$ and $z_{out}(r)$ match in the region 
$\sqrt{\epsilon} \ll r \ll 1$. The last matching condition
also sets $a=1$.

To complete our first order asymptotic perturbation analysis we have
to find $z_{equ}(r)$ for $r \in (1-\sqrt{\epsilon},1]$. We note that 
unlike the BC (\ref{cylsurfpbcz}), which gives rise to a new singularity
on top of the spherical profile $z_{out}(r)$ in the neck region, 
the divergence of the outer expansion (\ref{outerexpand}) 
for $1-r \sim \epsilon$, is caused simply because the original singularity,
$z^{\prime}(r) \to \infty$ at $r_s=1$, is shifted to some $r_s(\epsilon) < 1$. 
Analysis of equation~(\ref{cylsurfpz}) around the singular point shows
that $z_{equ}(r)$ satisfies:
\eq
z_{equ}^{\prime} = c(\epsilon)  / \sqrt{r_s(\epsilon) - r}  , \qquad 1-r \ll 1 , 
\label{equfunc}
\en  
where 
\eq
c(\epsilon) = 1 + O(\epsilon) , \qquad r_s(\epsilon) = 1 + O(\epsilon)  . 
\label{equexpand}
\en
The functions $c(\epsilon)$ and $r_s(\epsilon)$ are determined from matching 
$z_{equ}$ to $z_{out}$ to $O(\epsilon)$ in the interval
$\sqrt{\epsilon} \ll  1-r \ll 1$. Performing the matching procedure 
yields
\eq
z_{equ}^{\prime} = \frac{1+\frac{1}{8} \epsilon}{\sqrt{2(1-\epsilon-r)}} + 
O(\epsilon^2)   .
\label{uptofirsteq}
\en

To conclude, to first order in $\epsilon$, the derivative $z^{\prime}(r)$ 
in the interval $(\epsilon,1-\epsilon)$ is given by four 
different functional forms, equations~(\ref{uptofirstout},\ref{uptofirstinner},
\ref{midregion},\ref{uptofirsteq}), 
each valid in a different sub-interval of $(\epsilon,r_s(\epsilon))$.
These functions match along the overlaps between the sub-intervals.
In terms of the variable $r$, they read
\wide
\begin{eqnarray}
z_{in}^{\prime}(r) &=& \frac{\epsilon}{\sqrt{r^2 -\epsilon^2}} + 
\frac{{r^2}}{\sqrt{r^2-\epsilon^2}} + \cdots  ;\qquad
{\rm{inner  \ region}}  :
\epsilon < r \ll \sqrt{\epsilon}  
\label{eqprofile1}\\ 
z_{mid}^{\prime}(r) &=& 
r + \frac{\epsilon}{r} + \frac{1}{2}r^3 + \frac{1}{2} \frac{\epsilon^3}{r^3} +
\frac{3}{2} \epsilon^{3/2}  +\cdots  ; \qquad  
{\rm{middle \  region}}  :
\epsilon \ll r \ll 1 
\label{eqprofile2}\\  
z_{out}^{\prime} &=& 
\frac{r}{\sqrt{1-r^2}} + \epsilon \frac{1}{r (1-r^2)^{3/2}} + \cdots 
 ; \qquad
{\rm{outer  \ region}} : 
\sqrt{\epsilon} \ll r < \ll 1-\sqrt{\epsilon} 
\label{eqprofile3}\\
z_{equ}^{\prime} &=& \frac{1+\frac{1}{8}\epsilon}{\sqrt{2(1-r-\epsilon)}}
+\cdots
; \qquad 
{\rm{equatorial  \ region}} :  1-r \ll 1   .  
\label{eqprofile4}
\end{eqnarray}   
\wideend
\subsubsection{Longitudinal stretching}
In order to get the actual profile $z(r)$, one has to 
integrate equations~(\ref{eqprofile1}-\ref{eqprofile4}). 
Since the four regions overlap each other
along finite intervals, artificial limits of integration (within
the overlapped intervals) must be introduced. Thus, we integrate
(\ref{eqprofile1}) from $\epsilon$ to $\epsilon^{\alpha}$ 
($1/2  < \alpha <1$), (\ref{eqprofile2}) from 
$\epsilon^{\alpha}$ to $\epsilon^{\beta}$ ($0  < \beta <1/2$),
(\ref{eqprofile3}) from 
$\epsilon^{\beta}$ to $1-\epsilon^{\gamma}$ ($0  < \gamma <1/2$), 
and (\ref{eqprofile4}) from 
$1-\epsilon^{\gamma}$ to $1-\epsilon$.
As will be verified below, the values of $\alpha,\beta,\gamma$ do
not appear in the final expressions to leading order in $\epsilon$.

Performing the integrals of equations~(\ref{eqprofile1}-\ref{eqprofile4})
between the above limits, we find the leading contributions to be:
\begin{eqnarray}
I_1 &=& \int_{\epsilon}^{\epsilon^{\alpha}} z_{in}^{\prime} dr 
\approx 
(\alpha - 1) \ \epsilon \log \epsilon + \cdots
 \nonumber \\
I_2 &=& \int_{\epsilon^{\alpha}}^{\epsilon^{\beta}} z_{mid}^{\prime} dr 
\approx  
\frac{1}{2} \epsilon^{2 \beta}  + (\beta - \alpha) \ \epsilon \log \epsilon
+\cdots 
\nonumber \\
I_3 &=& \int_{\epsilon^{\beta}}^{1-\epsilon^{\gamma}} z_{out}^{\prime} dr 
\approx
1 -\sqrt{2} \epsilon^{\gamma/2} - 
\frac{1}{2} \epsilon^{2 \beta}
- \beta \ \epsilon \log \epsilon 
+\cdots
\nonumber \\
I_4 &=& \int_{1-\epsilon^{\gamma}}^{1-\epsilon} z_{equ}^{\prime} dr 
\approx
\sqrt{2} \epsilon^{\gamma / 2}
+\cdots
\label{eqnprofile} 
\end{eqnarray} 
In particular, the $z$-coordinates of the centers of the spheres (in practice, the 'equatorial' values of $z$ at which $dz/dr=0$) are shifted 
from $\pm 1$ (for $\epsilon=0$)
to $ \pm z_s = \pm z(r_s(\epsilon))$:
\eq
z_s =  I_1 + I_2 + I_3 + I_4 \approx 
 1 + 
\epsilon \vert \log \epsilon \vert .  
\label{newzs}
\en  
\subsection{Dynamics}
In a recent work \cite{porous-prl}
we proved that all CMC surfaces, except the plane, 
are unstable under the dynamics~(\ref{surdyn}). This holds in particular
for the fixed point surfaces described by equations~(\ref{eqprofile1}-\ref{eqprofile4}).
Consider a surface $\bx$ close to a fixed point surface $\bx^{*}$:
\begin{equation}
\bx (\bx^{*}) = \bx^{*} + \delta (\bx^{*}) \bn (\bx^{*}) , 
\label{surfacepert}
\end{equation}
where $\bn$ is a unit vector normal to the unperturbed surface, 
and $\delta$ is the magnitude of the perturbation.  
Linear stability analysis of the dynamics~(\ref{surdyn}) near 
$\bx^{*}$ results in the formal equation for 
the perturbation $\delta$:
\eq
\frac{\partial \delta}{\partial t} = \cL \{ \delta \} , 
\label{linearoperator}
\en
where $\cL$ is a linear differential, 
generally non-Hermitian, operator,
whose coefficients depend on the geometry of the surface $\bx^{*}$.
In this section we will compute, to leading order in $\epsilon$,
the largest eigenvalue $\lambda_{max} (\cL)$, 
for $\bx^{*}$ a CSCMC whose profile
$z(r)$ is described by equations~(\ref{eqprofile1}-\ref{eqprofile4}). $\lambda_{max}$
determines the rate of growth of a generic perturbation $\delta$ with a non-zero component
along the maximal eigenvalue of equation~(\ref{linearoperator}), whether
it leads to sintering of the spheres or to dissolution of the neck. 

Derivation of the detailed form of $\cL$ 
requires some tools of differential geometry, and will not be presented here.
For our purposes it is enough to use the following result 
\cite{porous-prl}:
{\it The eigenvalues of $\cL$ are identical
to the eigenvalues of a Hermitian operator $\cH$:}
\begin{equation}
\cH = \frac{1}{2} \nabla_s^2 + V(\bx^{*})   , 
\label{hermitianoperator}
\end{equation}      
where $\nabla_s$ is a surface gradient, and the `potential' $V(\bx^{*})$
is:
\begin{eqnarray}
V(\bx^{*}) &=& 2H_{*}^2 - K - \frac{1}{2} [\nabla_s^2 B + (\nabla_s B)^2 ] 
\nonumber \\
B(\bx^{*}) &\equiv& - \frac{1}{8} \log (H_{*}^2 - K(\bx^{*}))  , 
\label{potential}
\end{eqnarray}  
where $H_{*}, K=\kappa_a \kappa_l$ are respectively the mean and Gaussian curvatures of the 
unperturbed CMC surface. In particular,  
$\lambda_{max}(\cL) =\lambda_{max}(\cH)$.   
For a CSCMC with profile $z(r)$ 
we use the coordinate system $(\hat{r},\hat{z},\hat{\phi})$ to write:
\begin{eqnarray}
\nabla_s^2 &=& \frac{1}{1+{z^{\prime}}^2} \nabla_r^2 - 
\frac{z^{\prime}z^{\prime\prime}}{(1+{z^{\prime}}^2)^2} \nabla_r ,
\nonumber \\
K (r) &=& \frac{\zp\zdp}{r(1+\zp^2)^2}    . 
\label{identities}
\end{eqnarray}   
We note that for the CSCMC described by equations~(\ref{eqprofile1}-\ref{eqprofile4}),    
the Gaussian curvature $K(r)$ and its derivatives diverge toward
the neck ($r \to \epsilon$), as $\epsilon \to 0$. This can be easily 
understood by recalling that $K=\kappa_a \kappa_l$. The largest value
of the azimuthal curvature $\kappa_a =1/r$ is 
clearly achieved at the neck, where 
$\kappa_{a,neck} = 1/ \epsilon$. Since for this surface
$\kappa_l = 2 - \kappa_a \approx -1 / \epsilon$, we see that 
$K_{neck} \approx - 1/\epsilon^2$ as $\epsilon \to 0$. 
As one moves away from the neck to the spherically shaped periphery, 
$\kappa_a \to 1$, the Gaussian curvature $K \to 1$, 
and therefore $|V|$ decreases. Thus, the maximal eigenvalue
$\lambda_{max}(\cH)$ 
is determined by the region near the neck, 
and we can compute it by considering the operator ${\cH}$ in the 
inner region $\epsilon \leq r \ll \sqrt{\epsilon}$ only. 
According to equations~(\ref{uptofirstinner},\ref{identities}), the change
of variables: 
\begin{equation}
x = \sqrt{r^2 - \epsilon^2}  , 
\label{changevariables}
\end{equation}  
enables us to write $\nabla_s^2 = \partial_x^2$ in the inner region.
Some tedious algebra (see appendix) yields the potential form:
\begin{equation}
V(x) = \frac{3}{4}\frac{\epsilon^2 + \frac{1}{6} x^2}{(\epsilon^2 +x^2)^2} , 
\label{harmonic}
\end{equation}
in the inner region. 
With the $1d$ potential (\ref{harmonic}), the eigenvalue equation for 
${\cal H}$ is equivalent to a spheroidal wave equation \cite{abramowitz-stegun}. 
Numerical calculation of $\lambda_{max}(\cH)$ in the limit $\epsilon \to 0$
yielded $\lambda_{max}(\cH) = 0.282 \epsilon^{-2} + O(\epsilon^{-1})$.

We conclude that the profile given by 
equations~(\ref{eqprofile1}-\ref{eqprofile4}) is an unstable fixed point 
of the dynamics~(\ref{cylsurdyn2}). 
Typical perturbations will increase 
at a rate proportional to $\epsilon^{-2}$, and will 
lead either to sintering of the two spheres, or to dissolution of the neck.

This has some immediate consequences. Consider, for example, two spheres compressed 
against one another so that the distance between their centers is $d <2R$. A sinter 
formed around the contact zone between such spheres will generally grow, since the 
contact geometry should typically be on the growth side of the critical neck geometry. 
On the other hand, if two spheres become connected by a neck while their centers are at 
a distance $d > 2R$, the dynamics of the neck may cause it to either grow or evaporate, 
the boundary between these two behaviors being approximated by 
$d \approx 2R(1+\epsilon \vert \log \epsilon \vert)$. 


\begin{center}
\begin{figure}
\includegraphics[scale=0.45]{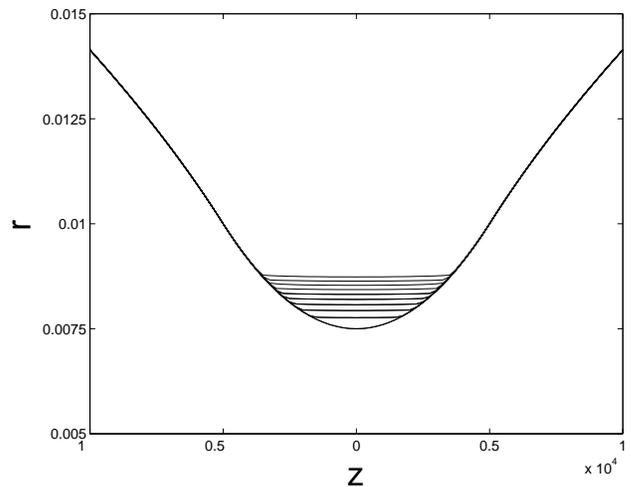}
\caption{Evolution of cylindrically symmetric sintering profile under 
equation~(\ref{cylsurdyn2}). The initial profile is parabolic.}
\label{figure3}
\end{figure}
\end{center}


\section{Asymptotic dynamics}

In this section, we will study the asymptotic behavior of
two sintering spheres when the surface 
profile is far from any fixed point, but the neck size is still small
compared to the radius of the spheres. 
In this regime, 
the non-linear dynamics (\ref{cylsurdyn2}) cannot be approximated by its 
linearized form (\ref{linearoperator}).
In figure 3 we show the numerically computed evolution of an asymptotic 
neck profile, which appears to evolve according 
to a similarity law. 

For volume preserving dynamics similar to 
(\ref{cylsurdyn2}), Kuczinsky \cite{kuczinsky49} suggested that the 
asymptotic neck profile 
evolves like a spherical cup,
\begin{equation}
r(z,t) = r_0 (t) + w(t)  - \sqrt{w^2 (t) - z^2}; 
\qquad 0 \leq z \leq w(t) , 
\label{kuzcinskicup}
\end{equation}       
where $r_0 (t)$ is the neck thickness and $w(t)$ its width. 
This neck profile has to be patched to the static peripheral spherical 
shape:
\begin{equation}
r(z,t) = \sqrt{2z - z^2}  .  
\label{peripherialsphere}
\end{equation}

Patching the $z$-coordinates of the two regions at $z=w(t)$ implies:
\eq
(w(t)  + r_0 (t))^2 \approx 2 w(t) \Rightarrow
w(t) \approx r_0 (t) ^2 / 2 \ll r_0 (t)  , 
\label{patchingz}
\en 
where we used $r_0 (t) \ll 1$.
To obtain the time dependence of $r_0 (t), w(t)$ 
we recall that the RHS of equation~(\ref{cylsurdyn2}) is proportional 
to $\frac{1}{2}(\kappa_a + \kappa_l)-1$, where
$\kappa_{a,neck} \approx 1/r_0 (t)$, and 
the spherical cup profile (\ref{kuzcinskicup}) implies
$\kappa_l \approx -1/w(t)$.  
According to equation~(\ref{patchingz}) $w(t) \ll r_0 (t) \ll 1$, and 
therefore the RHS of (\ref{cylsurdyn2}) is dominated by ${w(t)}^{-1}$,
whereas the LHS is simply ${d r_0 }/{dt}$. 
Using equation~(\ref{patchingz}) we get 
\eq
\frac{dr_0}{dt} \approx \frac{1}{r_0^2} \Rightarrow r_0 \sim t^{1/3}  , 
\label{timescaling}
\en
in agreement with experiments. A similar derivation for 
sintering dominated by surface diffusion yields $r_0 \sim t^{1/7}$,
and other growth rates can be derived for other kinds of kinetic 
routes \cite{kuczinsky49,herring50,swinkels81,german86}.
 
It should be stressed here that Kuczinsky's profile is simply a guess
and is not in any sense an asymptotic solution of equation~(\ref{cylsurdyn2})
or of its volume preserving version. 
Moreover, notice that the $t^{1/3}$ growth rate cannot be derived
for a general neck profile, but rather depends on the assumed shape.
As an example, consider a parabolic neck profile:
\eq
r(z,t) = r(t) + a(t) z^2   .   
\label{parabolic}
\en
Patching profiles and derivatives with the 
limit $z \to 0$ of (\ref{peripherialsphere}) gives $r_0 \sim t^{1/4}$.

Kuczinsky's assumption of semi-circular 
neck profile (\ref{kuzcinskicup}) was 
questioned by several authors, and other possible asymptotic neck 
profiles were suggested for sintering dominated 
by evaporation-precipitation and by other kinetic routes. 
In particular, several researchers 
\cite{german75,amar89},
conjectured that in a variety of sintering processes, the 
neck profile evolves as a CSCMC surface, 
whose spatially constant mean curvature
evolves in time. Notice the similarity between this assumption and Kuczinsky's 
one, according to which the longitudinal curvature $\kappa_l$ is 
spatially constant at the neck, whereas $\kappa_a$ may slightly vary.      
Other researchers, who focused their analysis 
on sintering dominated by surface diffusion, 
performed careful numerical simulations 
and advanced analytic arguments, from which they concluded 
that the asymptotic neck shape evolves 
very differently from the shape suggested by 
Kuczinsky, equation (\ref{kuzcinskicup}),
\cite{nichols65,german78,eggers98}. 
Moreover, it was shown by these researchers that 
in that case the asymptotic growth rate of the neck is not $t^{1/7}$,
as was suggested by Kuczinsky-like arguments for surface diffusion
dynamics, but rather becomes closer to $t^{1/6}$ or even $t^{1/5}$. 
Other studies, which focused on kinetics
dominated by surface and grain boundary diffusion showed that 
the neck profile and its growth rate in this case are 
also much more complicated than the simple models suggested by 
Kuczinsky \cite{bross79,swinkels80}.   

Here we show that the asymptotic neck profile for the sintering process
determined by equation~(\ref{cylsurdyn2}) evolves to a similarity shape, 
which is
indeed different from Kuczinsky's or other suggested shapes. 
We show however, that Kuczinsky's $t^{1/3}$ growth rate is correct in
this case. 

Let us write the sinter profile $r(z,t)$ as
\eq
r(z,t) = r_0 (t) + b(z,t) ,
\label{suggested}
\en
with $b(0)=0$, so that $r_0(t)$ is simply the radius at the center of the sinter. 
We suppose that throughout the sintering, non-spherical, region, we can assume
\eq
b(z,t) \ll r_0(t),
\en
an assumption we must verify for self-consistency at the conclusion of the 
calculation. Then we have
\eq
\frac{b^{\prime\prime}}{1+{b^{\prime}}^2} \approx 2\dot{r_0} 
 + \frac{1}{r_0} \equiv \alpha (t),
\label{dynsuggested} 
\en  
where we have replaced $r$ by $r_0$ and 
$2 \sqrt{1+{r^{\prime}}^2} \approx 0$, their values at $z=0$, 
while retaining the longitudinal curvature term (which also dominates the curvature in Kuczinsky's solution).
The solution of equation~(\ref{dynsuggested}) is:
\begin{equation}
b(z,t) \approx - \frac{1}{\alpha (t)} \log \cos [\alpha (t) z] ,
\label{solsuggested}
\end{equation}      
where we assumed the boundary condition: $b^{\prime}(0) = 0$ 
(from $z \leftrightarrow -z$ symmetry).

Following a similar reasoning to the one 
presented in the beginning of this section, 
we notice that patching derivatives of the concave neck profile given by 
equations~(\ref{suggested},\ref{solsuggested}), and the 
convex spherical periphery, equation~(\ref{peripherialsphere}) is possible 
only for $(r,z) \to (r^{*},z^{*})$ where both derivatives become 
much larger than one:
\eq
\left.\frac{\partial r}{\partial z}\right\vert_{z^*} 
\approx \tan [\alpha (t) \ z^{*}] \approx \frac{1}{\pi /2 
-  \alpha (t) z^{*}} \approx \frac{1}{\sqrt{2 z^{*}}} ,
\label{patchingsingular}
\en
from which we get:
\eq
z^{*} = \frac{\pi}{2 \alpha (t)} + O(\alpha^{-3/2}).
\label{fromwhich}
\en
Matching to the peripheral spherical shape equation (\ref{peripherialsphere})
implies:
\eq
r^{*} \approx \sqrt{2 z^{*}} .
\label{fromwhich2} 
\en
From equations~(\ref{dynsuggested},\ref{fromwhich},\ref{fromwhich2}) we get:
\eq
2\dot{r_0} \approx -\frac{1}{r_0} + \frac{\pi}{r_0^2} \approx 
\frac{\pi}{r_0^2} , 
\en
whose asymptotic solution [at $r_0(t=0) \ll r_0 (t) \ll 1$] 
gives the Kuczinsky growth rate (\ref{timescaling}),
\eq
r_0(t) \approx \left(\frac{3 \pi t}{2}\right)^{1/3}   .
\en

To check the consistency of this solution we must verify 
the validity of the approximation $\vert b(z,t)/r_0(t) \vert \ll 1$ for $z < z^{*}$. 
This seems clear, since although
$b^{\prime}$ is large at the patching point for small $z^{*}$, $b$ is still 
small ($b \ll r_0$) because of equations~(\ref{solsuggested},\ref{fromwhich}).
However, while inside the neck, near $z=0$, the 
term $2\sqrt{1+{{z^{\prime}}^2}}$ (excluded from equation~(\ref{dynsuggested}))is negligible,
we should insure that it does not become large as
$z \to z^{*}$. Therefore this term should be included in the consistency check,
and we seek a solution to the equation:
\begin{equation} 
\frac{b^{\prime\prime}}{1+{b^{\prime}}^2}  + 2 \sqrt{1 + {b^{\prime}}^2} = 2\dot{r_0} 
 + \frac{1}{r_0} = 
\alpha (t), 
\label{newmatch}
\end{equation}
under the same BC as above: $b(0)=b^{\prime}(0)=0$. 
With the change of variables: 
$\tan u = b^{\prime}$, equation~(\ref{newmatch}) recasts into the form:
\begin{equation} 
u^{\prime} + 2\sqrt{1+\tan^2 u}= 
\alpha (t)  ,
\label{newmatch2}
\end{equation}
from which we get
\begin{eqnarray}
b(z^{*}) &=& \int_0^{z^{*}} \!\!\! dz\ b^{\prime} = \int_0^{\tan^{-1}[b^{\prime}(z^*)]} 
du \frac{\tan u}{\alpha (t) - 2\sqrt{1 + \tan^2 u}}\nonumber \\
&\approx& \frac{1}{2 \alpha (t)} \log \left(1 + [b^{\prime}(z^{*})]^2 \right).
\label{newmatch3}
\end{eqnarray}
This result is identical for small $z^{*}$ to equation~(\ref{solsuggested}),
from which we verify the consistency of equation~(\ref{patchingsingular}).     
 
\section{Conclusions}
 
In this paper we studied a sintering 
model that we believe to be 
relevant for understanding
the evolution of morphologies in sedimentary rocks  
and other geophysical systems,
under reaction-limited kinetics and open environmental conditions. The assumption 
of reaction as opposed to transport limited kinetics is quite defensible in this context. 
For example, for siliciclastic rocks 
(quartz in water) we take as representative values:
$L_{{\rm par}} \approx L_{{\rm pore}} \approx 10^{-4}\rm{m}$,
$D \approx 10^{-9} \rm{m^2/s}$,
and $K_f \approx 2 \ 10^{-15}\rm{m/s}$. From equation~(\ref{connecttimes})
we obtain the typical time scales
$\tau_{{\rm diss}} \approx 10^{11} \rm{s} \gg \tau_{{\rm diff}} \approx 10 \rm{s}$. 
However, much of the diagenesis of sedimentary rocks is driven by pressure solution, 
rather than by surface energy effects \cite{pressure-solution,pressure-rutter}. 
Surface energy effects might dominate in systems in which the solid phase has not 
experienced significant pressure changes over the lifetime of the grains--a situation that might prevail in magmatic plumes, in which solidified material coexists 
with liquid magma still saturated with the components of this material \cite{visser99}.

In glaciology, the most likely complication to our assumption of dissolution-precipitation 
kinetics driven by surface energies is likely to be surface melting, especially near the 
melting temperature of the ice. Transport via surface diffusion or in melted surface 
layers is also likely to be significant for sintered metallic powders, especially if 
the vapor pressure of the metals is low.

Even though the sintering physics of real technological and geological materials is 
likely rarely to be in a regime where the rather simplified model of this work includes 
all of the relevant physics, we nevertheless believe that our methods, based on careful 
asymptotic analysis of the stationary states of the kinetics, combined with similarity 
analysis of late-stage sintering, not only illuminate the physical limit we have chosen, 
but also should deal successfully with other types of tranport mechanisms. This is 
likely to be the subject of subsequent works.

\bibliography{sinter}

\begin{thebibliography}{27}
\expandafter\ifx\csname natexlab\endcsname\relax\def\natexlab#1{#1}\fi
\expandafter\ifx\csname bibnamefont\endcsname\relax
  \def\bibnamefont#1{#1}\fi
\expandafter\ifx\csname bibfnamefont\endcsname\relax
  \def\bibfnamefont#1{#1}\fi
\expandafter\ifx\csname citenamefont\endcsname\relax
  \def\citenamefont#1{#1}\fi
\expandafter\ifx\csname url\endcsname\relax
  \def\url#1{\texttt{#1}}\fi
\expandafter\ifx\csname urlprefix\endcsname\relax\def\urlprefix{URL }\fi
\providecommand{\bibinfo}[2]{#2}
\providecommand{\eprint}[2][]{\url{#2}}

\bibitem[{\citenamefont{Abramowitz and Stegun}(1970)}]{abramowitz-stegun}
\bibinfo{author}{\bibnamefont{Abramowitz}, \bibfnamefont{M.}}, and
  \bibinfo{author}{\bibfnamefont{I.}~\bibnamefont{Stegun}},
  \bibinfo{year}{1970}, \emph{\bibinfo{title}{Handbook of Mathematical
  Functions}} (\bibinfo{publisher}{Dover}, \bibinfo{address}{New York}).

\bibitem[{\citenamefont{Amar} \emph{et~al.}(1989)\citenamefont{Amar, Bernholc,
  Berry, Jellinek, and Salamon}}]{amar89}
\bibinfo{author}{\bibnamefont{Amar}, \bibfnamefont{F.}},
  \bibinfo{author}{\bibfnamefont{J.}~\bibnamefont{Bernholc}},
  \bibinfo{author}{\bibfnamefont{R.}~\bibnamefont{Berry}},
  \bibinfo{author}{\bibfnamefont{J.}~\bibnamefont{Jellinek}}, and
  \bibinfo{author}{\bibfnamefont{P.}~\bibnamefont{Salamon}},
  \bibinfo{year}{1989}, \bibinfo{journal}{J. Appl. Phys.}
  \textbf{\bibinfo{volume}{65}}, \bibinfo{pages}{3219}.

\bibitem[{\citenamefont{de~Boer}(1977)}]{pressure-solution}
\bibinfo{author}{\bibnamefont{de~Boer}, \bibfnamefont{R.}},
  \bibinfo{year}{1977}, \bibinfo{journal}{Geochim. Cosmochim. Acta}
  \textbf{\bibinfo{volume}{41}}, \bibinfo{pages}{249}.

\bibitem[{\citenamefont{Bray}(1994)}]{bray94}
\bibinfo{author}{\bibnamefont{Bray}, \bibfnamefont{A.}}, \bibinfo{year}{1994},
  \bibinfo{journal}{Adv. Phys.} \textbf{\bibinfo{volume}{43}},
  \bibinfo{pages}{357}.

\bibitem[{\citenamefont{Bross and Exxner}(1979)}]{bross79}
\bibinfo{author}{\bibnamefont{Bross}, \bibfnamefont{P.}}, and
  \bibinfo{author}{\bibfnamefont{H.}~\bibnamefont{Exxner}},
  \bibinfo{year}{1979}, \bibinfo{journal}{Acta Metall.}
  \textbf{\bibinfo{volume}{27}}, \bibinfo{pages}{1013}.

\bibitem[{\citenamefont{Colbeck}(1998)}]{colbeck98}
\bibinfo{author}{\bibnamefont{Colbeck}, \bibfnamefont{S.}},
  \bibinfo{year}{1998}, \bibinfo{journal}{J. Appl. Phys.}
  \textbf{\bibinfo{volume}{84}}, \bibinfo{pages}{4585}.

\bibitem[{\citenamefont{Davidovitch}
  \emph{et~al.}(2002)\citenamefont{Davidovitch, Erta{\c s}, and
  Halsey}}]{porous-prl}
\bibinfo{author}{\bibnamefont{Davidovitch}, \bibfnamefont{B.}},
  \bibinfo{author}{\bibfnamefont{D.}~\bibnamefont{Erta{\c s}}}, and
  \bibinfo{author}{\bibfnamefont{T.}~\bibnamefont{Halsey}},
  \bibinfo{year}{2002}, \bibinfo{howpublished}{cond-mat/0210503}.

\bibitem[{\citenamefont{Eggers}(1998)}]{eggers98}
\bibinfo{author}{\bibnamefont{Eggers}, \bibfnamefont{J.}},
  \bibinfo{year}{1998}, \bibinfo{journal}{\prl} \textbf{\bibinfo{volume}{80}},
  \bibinfo{pages}{2634}.

\bibitem[{\citenamefont{German}(1986)}]{german86}
\bibinfo{author}{\bibnamefont{German}, \bibfnamefont{R.}},
  \bibinfo{year}{1986}, \emph{\bibinfo{title}{Liquid Phase Sintering}}
  (\bibinfo{publisher}{Plenum Publishing}, \bibinfo{address}{New York}).

\bibitem[{\citenamefont{German}(1996)}]{german96}
\bibinfo{author}{\bibnamefont{German}, \bibfnamefont{R.}},
  \bibinfo{year}{1996}, \emph{\bibinfo{title}{Sintering Theory and Practice}}
  (\bibinfo{publisher}{John Wiley and Sons}, \bibinfo{address}{New York}),
  \bibinfo{edition}{1st} edition.

\bibitem[{\citenamefont{German and Lathrop}(1978)}]{german78}
\bibinfo{author}{\bibnamefont{German}, \bibfnamefont{R.}}, and
  \bibinfo{author}{\bibfnamefont{J.}~\bibnamefont{Lathrop}},
  \bibinfo{year}{1978}, \bibinfo{journal}{J. Mater. Sci.}
  \textbf{\bibinfo{volume}{13}}, \bibinfo{pages}{921}.

\bibitem[{\citenamefont{German and Munir}(1975)}]{german75}
\bibinfo{author}{\bibnamefont{German}, \bibfnamefont{R.}}, and
  \bibinfo{author}{\bibfnamefont{Z.}~\bibnamefont{Munir}},
  \bibinfo{year}{1975}, \bibinfo{journal}{Metal. Trans. B}
  \textbf{\bibinfo{volume}{6B}}, \bibinfo{pages}{289}.

\bibitem[{\citenamefont{Hay and Evans}(1988)}]{hay88}
\bibinfo{author}{\bibnamefont{Hay}, \bibfnamefont{R.}}, and
  \bibinfo{author}{\bibfnamefont{B.}~\bibnamefont{Evans}},
  \bibinfo{year}{1988}, \bibinfo{journal}{J. Geophys. Res.}
  \textbf{\bibinfo{volume}{93}}, \bibinfo{pages}{8959}.

\bibitem[{\citenamefont{Herring}(1950)}]{herring50}
\bibinfo{author}{\bibnamefont{Herring}, \bibfnamefont{C.}},
  \bibinfo{year}{1950}, \bibinfo{journal}{J. Appl. Phys.}
  \textbf{\bibinfo{volume}{21}}, \bibinfo{pages}{301}.

\bibitem[{\citenamefont{Hinch}(1991)}]{hinch91}
\bibinfo{author}{\bibnamefont{Hinch}, \bibfnamefont{E.}}, \bibinfo{year}{1991},
  \emph{\bibinfo{title}{Perturbation Methods}} (\bibinfo{publisher}{Cambridge
  University Press}, \bibinfo{address}{Cambridge}).

\bibitem[{\citenamefont{Hobbs and Mason}(1964)}]{hobbs64}
\bibinfo{author}{\bibnamefont{Hobbs}, \bibfnamefont{P.}}, and
  \bibinfo{author}{\bibfnamefont{B.}~\bibnamefont{Mason}},
  \bibinfo{year}{1964}, \bibinfo{journal}{Phil. Mag.}
  \textbf{\bibinfo{volume}{8}}, \bibinfo{pages}{181}.

\bibitem[{\citenamefont{Jurewicz and Watson}(1985)}]{jurewicz85}
\bibinfo{author}{\bibnamefont{Jurewicz}, \bibfnamefont{S.}}, and
  \bibinfo{author}{\bibfnamefont{E.}~\bibnamefont{Watson}},
  \bibinfo{year}{1985}, \bibinfo{journal}{Geochim. Cosmochim. Acta}
  \textbf{\bibinfo{volume}{49}}, \bibinfo{pages}{1109}.

\bibitem[{\citenamefont{Kingery}(1960)}]{kingery60}
\bibinfo{author}{\bibnamefont{Kingery}, \bibfnamefont{W.}},
  \bibinfo{year}{1960}, \bibinfo{journal}{J. Appl. Phys.}
  \textbf{\bibinfo{volume}{31}}, \bibinfo{pages}{833}.

\bibitem[{\citenamefont{Kingery and Berg}(1955)}]{kingery55}
\bibinfo{author}{\bibnamefont{Kingery}, \bibfnamefont{W.}}, and
  \bibinfo{author}{\bibfnamefont{M.}~\bibnamefont{Berg}}, \bibinfo{year}{1955},
  \bibinfo{journal}{J. Appl. Phys.} \textbf{\bibinfo{volume}{26}},
  \bibinfo{pages}{1205}.

\bibitem[{\citenamefont{Kuczinsky}(1949)}]{kuczinsky49}
\bibinfo{author}{\bibnamefont{Kuczinsky}, \bibfnamefont{G.}},
  \bibinfo{year}{1949}, \bibinfo{journal}{J. Metal.}
  \textbf{\bibinfo{volume}{1}}, \bibinfo{pages}{169}.

\bibitem[{\citenamefont{Kuroiwa}(1961)}]{kuroiwa61}
\bibinfo{author}{\bibnamefont{Kuroiwa}, \bibfnamefont{D.}},
  \bibinfo{year}{1961}, \bibinfo{journal}{Tellus}
  \textbf{\bibinfo{volume}{12}}, \bibinfo{pages}{252}.

\bibitem[{\citenamefont{Maeono and Ebinuma}(1983)}]{maono83}
\bibinfo{author}{\bibnamefont{Maeono}, \bibfnamefont{N.}}, and
  \bibinfo{author}{\bibfnamefont{T.}~\bibnamefont{Ebinuma}},
  \bibinfo{year}{1983}, \bibinfo{journal}{J. Phys. Chem.}
  \textbf{\bibinfo{volume}{87}}, \bibinfo{pages}{4103}.

\bibitem[{\citenamefont{Nichols and Mullins}(1965)}]{nichols65}
\bibinfo{author}{\bibnamefont{Nichols}, \bibfnamefont{F.}}, and
  \bibinfo{author}{\bibfnamefont{W.}~\bibnamefont{Mullins}},
  \bibinfo{year}{1965}, \bibinfo{journal}{J. Appl. Phys.}
  \textbf{\bibinfo{volume}{36}}, \bibinfo{pages}{1826}.

\bibitem[{\citenamefont{Rutter}(1976)}]{pressure-rutter}
\bibinfo{author}{\bibnamefont{Rutter}, \bibfnamefont{E.~H.}},
  \bibinfo{year}{1976}, \bibinfo{journal}{Phil. Trans. R. Soc. Lond. A}
  \textbf{\bibinfo{volume}{283}}, \bibinfo{pages}{203}.

\bibitem[{\citenamefont{Swinkels and Ashby}(1980)}]{swinkels80}
\bibinfo{author}{\bibnamefont{Swinkels}, \bibfnamefont{F.}}, and
  \bibinfo{author}{\bibfnamefont{M.}~\bibnamefont{Ashby}},
  \bibinfo{year}{1980}, \bibinfo{journal}{Powder Metall.}
  \textbf{\bibinfo{volume}{1}}, \bibinfo{pages}{1}.

\bibitem[{\citenamefont{Swinkels and Ashby}(1981)}]{swinkels81}
\bibinfo{author}{\bibnamefont{Swinkels}, \bibfnamefont{F.}}, and
  \bibinfo{author}{\bibfnamefont{M.}~\bibnamefont{Ashby}},
  \bibinfo{year}{1981}, \bibinfo{journal}{Powder Metall.}
  \textbf{\bibinfo{volume}{29}}, \bibinfo{pages}{259}.

\bibitem[{\citenamefont{Visser}(1999)}]{visser99}
\bibinfo{author}{\bibnamefont{Visser}, \bibfnamefont{H.}},
  \bibinfo{year}{1999}, \emph{\bibinfo{title}{Mass transfer processes in
  crystalline aggregates containing a fluid phase}}, Ph.D. thesis,
  \bibinfo{school}{Utrecht University}.

\end{thebibliography}

\section*{APPENDIX}

In this appendix we prove that the potential $V$, as defined in 
equation~(\ref{potential}) has the form~(\ref{harmonic}) in the inner region 
$z_{in}(r)$. For the CSCMC surface~(\ref{eqprofile1}-\ref{eqprofile4}) $H=1$, $\nabla_s H =0$, and 
from equation~(\ref{potential}) we get:
\begin{equation}
V = 2 - K - \frac{1}{16} \frac{\nabla_s^2 K}{1-K} - 
\frac{9}{128}\frac{(\nabla_s K)^2}{(1-K)^2} \ . 
\label{appen1}
\end{equation}
From equation~(\ref{identities}) we get:
\wide
\begin{eqnarray}
|\nabla_s K| &=& \frac{1}{r(1+\zp^{2})^{5/2}}
\ \ [  \ \ \zdp^2 + \zp \ztp - \frac{\zp \zdp}{r} - 
\frac{4\zp^2 \zdp^2}{1 + \zp^2} \ \ ] \nonumber  \\
{\nabla_s}^2 K &=& \frac{1}{r (1+\zp^2)^3} 
\ \ [ \ \ 3\zdp \ztp + \zp \zqp + 2\frac{\zp \zdp}{r^2} 
-2 \frac{\zp\ztp}{r}
- 2\frac{{\zdp}^2}{r}
\nonumber \\
&-& 13 \frac{\zp{\zdp}^3}{1+\zp^{2}}
-13 \frac{{\zp}^2\zdp \ztp}{1+\zp^{2}}
+28 \frac{{\zp}^3{\zdp}^3}{(1+\zp^{2})^2}
+9 \frac{{\zp}^2{\zdp}^2}{r(1+\zp^{2})} ] . 
\label{appen2}
\end{eqnarray}
\wideend
Consider now the profile $z_{in}(r)$ in the inner region
$\epsilon \leq r \ll \sqrt{\epsilon}$, as given in 
equation~(\ref{eqprofile1}). Since in this region 
$r^2 / (r^2 - \epsilon^2) \ll 1$, we find that the leading contributions 
to the profile derivatives are:
\begin{eqnarray}
\zp = \frac{\epsilon}{(r^2-\epsilon^2)^{1/2}},  &\qquad&
\zdp = -\frac{\epsilon r}{(r^2-\epsilon^2)^{3/2}}, \nonumber \\
\ztp = \frac{2\epsilon r^2+\epsilon^3}{(r^2-\epsilon^2)^{5/2}}, &\qquad&
\zqp = -\frac{6\epsilon r^3-9r\epsilon^3}{(r^2-\epsilon^2)^{7/2}} . 
\label{appen3}
\end{eqnarray}  
Substituting the last expressions for the profile derivatives
in equation~({\ref{appen2}), and using the change of 
variables~(\ref{changevariables}), we obtain equation~(\ref{harmonic}).

\end{document}
